\documentclass[aip,reprint]{revtex4-1}

\usepackage{graphicx}
\graphicspath{{figures/}}
\usepackage{xspace}
\usepackage{hyperref}
\newcommand{\kalpha}{K\ensuremath{\alpha}\xspace}
\newcommand{\mnka}{Mn \kalpha}
\newcommand{\crka}{Cr \kalpha}
\newcommand{\kaone}{K\ensuremath{\alpha_1}\xspace}
\newcommand{\katwo}{K\ensuremath{\alpha_2}\xspace}
\newcommand{\crkaone}{Cr \kaone}
\newcommand{\cukaone}{Cu \kaone}
\newcommand{\fefiftyfive}{\ensuremath{^{55}}Fe\xspace}
\newcommand{\procspie}{Proc. SPIE}

\begin{document}


\title{Simple, compact, high-resolution monochromatic x-ray source
  for characterization of x-ray calorimeter arrays} 



\author{M.~A.~Leutenegger} 
\email{maurice.a.leutenegger@nasa.gov}
\affiliation{Code 662, NASA Goddard Space Flight Center, Greenbelt, MD
  20771, USA}
\author{M.~E.~Eckart} 
\affiliation{Lawrence Livermore National Laboratory, Livermore, CA 94550, USA}
\author{S.~J.~Moseley} 
\affiliation{Code 662, NASA Goddard Space Flight Center, Greenbelt, MD
  20771, USA}
\author{S.~O.~Rohrbach} 
\affiliation{Code 551, NASA Goddard Space Flight Center, Greenbelt, MD
  20771, USA}
\author{J.~K.~Black}
\affiliation{Rock Creek Scientific, 1400 East-West Hwy, Suite 807, Silver Spring, MD 20910, USA}
\author{M.~P.~Chiao} 
\affiliation{Code 592, NASA Goddard Space Flight Center, Greenbelt, MD
  20771, USA}
\author{R.~L.~Kelley} 
\affiliation{Code 662, NASA Goddard Space Flight Center, Greenbelt, MD
  20771, USA}
\author{C.~A.~Kilbourne}
\affiliation{Code 662, NASA Goddard Space Flight Center, Greenbelt, MD
  20771, USA}
\author{F.~S.~Porter}
\affiliation{Code 662, NASA Goddard Space Flight Center, Greenbelt, MD
  20771, USA}




\begin{abstract}
    X-ray calorimeters routinely achieve very high spectral resolution, typically a few eV full width at half maximum (FWHM). Measurements of calorimeter line shapes are usually dominated by the natural linewidth of most laboratory calibration sources.  This compounds the data acquisition time necessary to statistically sample the instrumental line broadening, and can add systematic uncertainty if the intrinsic line shape of the source is not well known. To address these issues, we have built a simple, compact monochromatic x-ray source using channel cut crystals. A commercial x-ray tube illuminates a pair of channel cut crystals which are aligned in a dispersive configuration to select the \kaone line of the x-ray tube anode material. The entire device, including x-ray tube, can be easily hand carried by one person and may be positioned manually or using a mechanical translation stage. The output monochromatic beam provides a collimated image of the anode spot with magnification of unity in the dispersion direction (typically 100-200 $\mu$m for the x-ray tubes used here), and is unfocused in the cross-dispersion direction, so that the source image in the detector plane appears as a line. We measured output count rates as high as 10 count/s/pixel for the Hitomi Soft X-ray Spectrometer, which had 819 $\mu$m square pixels. We implemented different monochromator designs for energies of 5.4 keV (one design) and 8.0 keV (two designs) which have effective theoretical FWHM energy resolution of 0.125, 0.197, and 0.086 eV, respectively; these are well-suited for optimal calibration measurements of state-of-the art x-ray calorimeters. We measured an upper limit for the energy resolution of our \crkaone monochromator of 0.7 eV FWHM at 5.4 keV, consistent with the theoretical prediction of 0.125 eV.
\end{abstract}

\pacs{}

\maketitle 

\section{Introduction}

X-ray calorimeters have been developed over the last 30 years with the aim of providing high-resolution, high-efficiency imaging spectroscopy for astrophysics and other applications \cite{1984JAP....56.1257M}. The Hitomi Soft X-ray Spectrometer (SXS), a production space flight spectrometer with a 36 pixel array of Si thermistor sensors with HgTe x-ray absorbers, achieved better than 5 eV full width at half maximum (FWHM) resolution at 6 keV in ground testing\cite{2018JATIS...4b1406E} and 5 eV on orbit\cite{2018JATIS...4b1407L}, and detectors based on Transition Edge Sensors (TES) as well as Metallic Magnetic  Calorimeters (MMC) have achieved better than 2 eV resolution at 6 keV and better than 1 eV below 2 keV\cite{2012JLTP..167..168S, 2014JLTP..176..617P, 2015ApPhL.107v3503L, 2017RScI...88e3108D, Kempf2018, 2019ITAS...2904472D}. Typical calorimeter pixels range in size from of order 0.01 to 1 mm$^2$, depending on the application.

The spectral response of x-ray calorimeter detectors is dominated by a Gaussian broadening that is often characterized using \mnka x-rays from a radioactive \fefiftyfive source, which has a large advantage in terms of ease of use.  The measured spectrum is a convolution of the intrinsic line shape of \mnka with the Gaussian core response of the detector. The intrinsic line shape of \mnka is highly complex, and is typically modeled using a 8 Lorentzian empirical deconvolution\cite{1997PhRvA..56.4554H, HolzerPrivComm}. Since the Lorentzian components have line widths of $\sim$ 2 eV, the intrinsic width of the complex is a significant fraction of the measured broadening for state-of-the-art calorimeters. This adds considerably to the integration time required to measure the intrinsic line width to the desired statistical precision given a fixed incident flux. This loss of experimental efficiency could be recovered with a sufficiently monochromatic x-ray source, which would return the maximum statistical precision achievable for the number of counts collected. Furthermore, characterization of non-Gaussian components of the detector response (from fluorescent escape photons, electron loss, and effects of incomplete thermalization)\cite{2018JATIS...4b1406E, 2019ITAS...2903420E} also requires a monochromatic x-ray source to avoid confusion between the intrinsic source spectrum and non-Gaussian components.

Thus, there is a clear need for monochromatic sources for characterization of x-ray calorimeters. High performance monochromators for experiments at synchrotron facilities are well-developed, but travel to such facilities is prohibitive for the day-to-day activities of a cryogenic physics laboratory. Thus we have built a series of small, portable monochromators using commercial table-top x-ray generators combined with channel cut crystals, and we have used these monochromators to characterize a number of x-ray calorimeter devices.

\section{Design}

\begin{figure}[ht]
  \includegraphics[width=84mm]{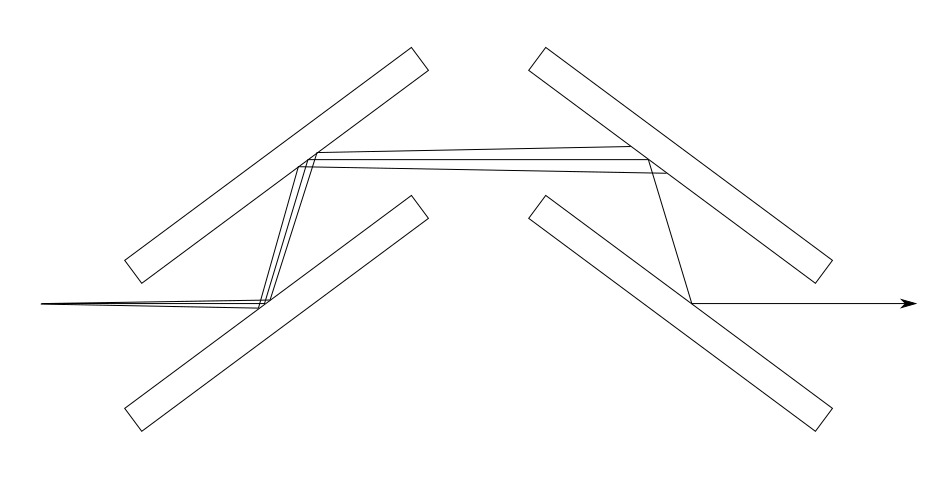}
  \caption{Schematic diagram of a DuMond-Hart-Bartels monochromator illuminated by a divergent broad-band point source. Rays of different energies satisfy the Bragg condition at different points on the first crystal pair, but only one ray satisfies the Bragg condition at the second crystal pair. (Actually the angular width of the reflected bundle has a small but finite width. See Section \ref{sec:theory}.)}
  \label{fig:dhb}
\end{figure}

The scheme used in our monochromators is due to DuMond, who first proposed a four reflection monochromator in the (+1,-1,-1,+1) configuration\cite{PhysRev.52.872}, as illustrated in Figure~\ref{fig:dhb}, although it was not implemented at the time due to insufficient flux produced by x-ray generators. The main advantage of such a multiple reflection scheme is the suppression of reflectivity in the wings of the crystal rocking curve. A secondary advantage is the preservation of the original beam propagation direction. This design also has the interesting property that the output beam produces a collimated image of the x-ray generator spot in the dispersion direction with magnification of unity.

\begin{figure*}[ht]
    \includegraphics[width=168mm]{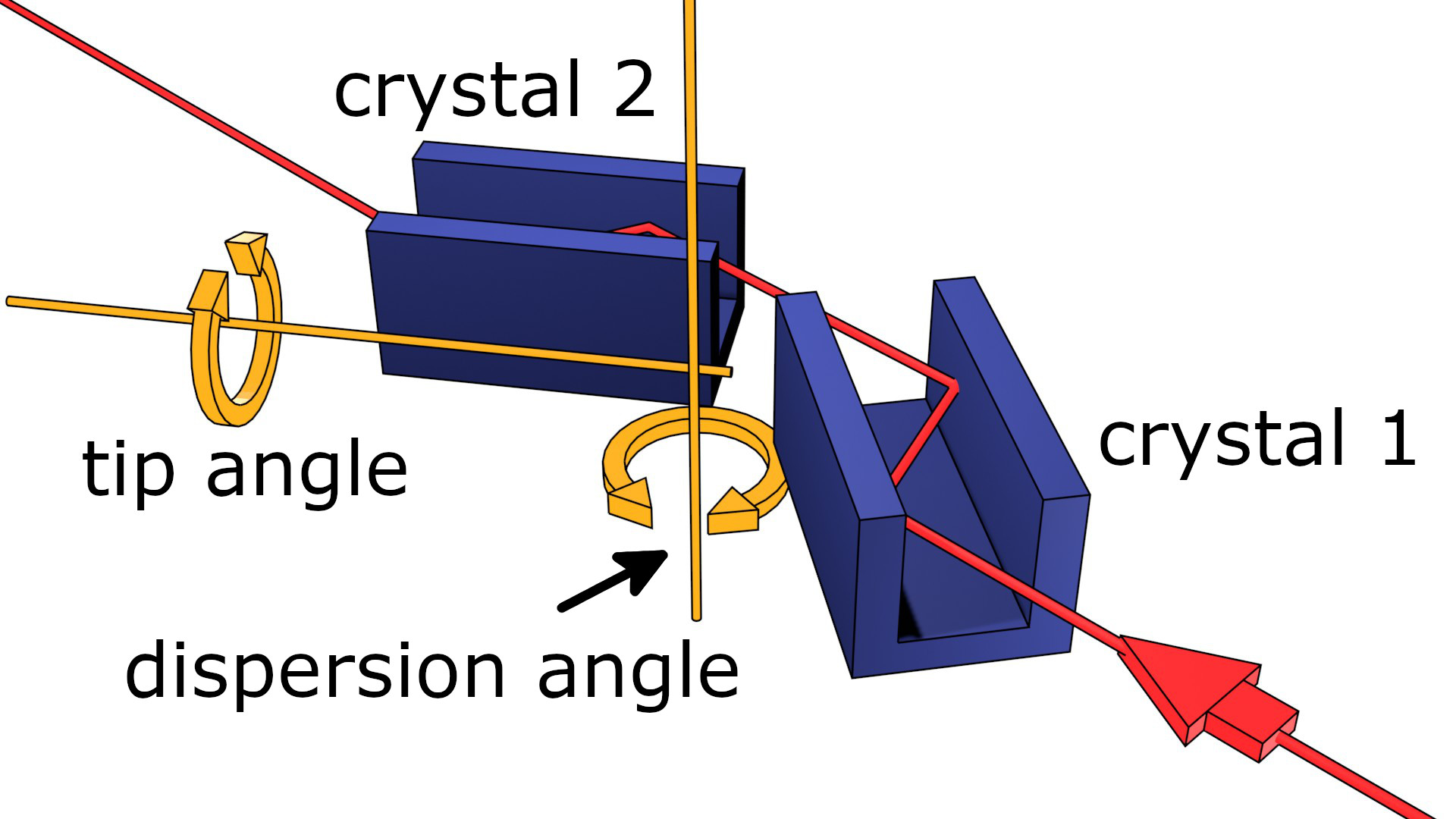}
    \caption{Visualization of our implementation for a 2+2 reflection DHB monochromator using two channel cut crystals (blue). The beam path is indicated (red), as are the axes for adjusting the dispersion angle and tip angle of the second crystal (yellow). Note that adjusting the dispersion angle of the second crystal by $\theta$ is equivalent to an adjustment of both crystals by $\theta/2$ with respect to an imaginary plane of symmetry; this plane also rotates by $\theta/2$ with respect to the lab frame. }
    \label{fig:cccm_diagram}
\end{figure*}

Subsequently Beaumont \& Hart and Bartels designed four reflection monochromators for synchrotron beams \cite{1974JPhE....7..823B, 1983JVSTB...1..338B}. Hart simplified the alignment problem by using channels cut in monolithic crystals to provide pairs of aligned reflecting surfaces. The most important remaining degree of freedom in the alignment is the angle between the two channel cut crystals in the dispersion plane, which determines the energy passed by the monochromator.

Such designs have been commonly implemented to take advantage of the bright x-rays of synchrotron beamlines; however, because x-ray calorimeter arrays are typically designed to handle only relatively low count rates (e.g. $\sim 1$ count/s/pixel for the Hitomi SXS), we can achieve acceptable output flux using a commercial air-cooled 50~W x-ray generator. 

To unambiguously characterize the performance of x-ray calorimeters, it is desirable to have a very monochromatic source with no flux in the line wings. Even a four reflection monochromator design achieves this, although marginal decreases in FWHM can be achieved with more reflections, at the cost of some throughput. The number of reflections in a given crystal design can be chosen by changing the ratio of channel length to channel width for a given Bragg angle such that 
\begin{equation}
    l = \frac{2Nw}{\tan \theta_b}\, ,
    \label{eqn:crystal_length}
\end{equation}
where $l$ is the channel length, $w$ is the channel width, $\theta_b$ is the Bragg angle, and $2N$ is the number of reflections per crystal in the design, with $N$ constrained to be an integer. In this article, we use the notation $2N+2N$ to denote the total number of reflections while calling attention to the symmetry in the optical design. All of our design implementations discussed in this article are for 4+4 or 6+6 reflection systems. 

We designed the monochromators to be simple and compact, using commercially available parts to the extent possible. All alignment stages are operated by hand, and the number of degrees of freedom has been minimized to five: cross dispersion translation and dispersion angle for each crystal; and the relative tip angle of the two crystals. In Fig.~\ref{fig:cccm_diagram}, we show an illustration of a 2+2 reflection channel cut crystal monochromator concept, indicating the path of the beam, as well as the axes for adjustment of the dispersion angle and relative tip angle of the second (downstream) crystal.

\section{Implementation}

\begin{figure*}[ht]
    \includegraphics[width=168mm]{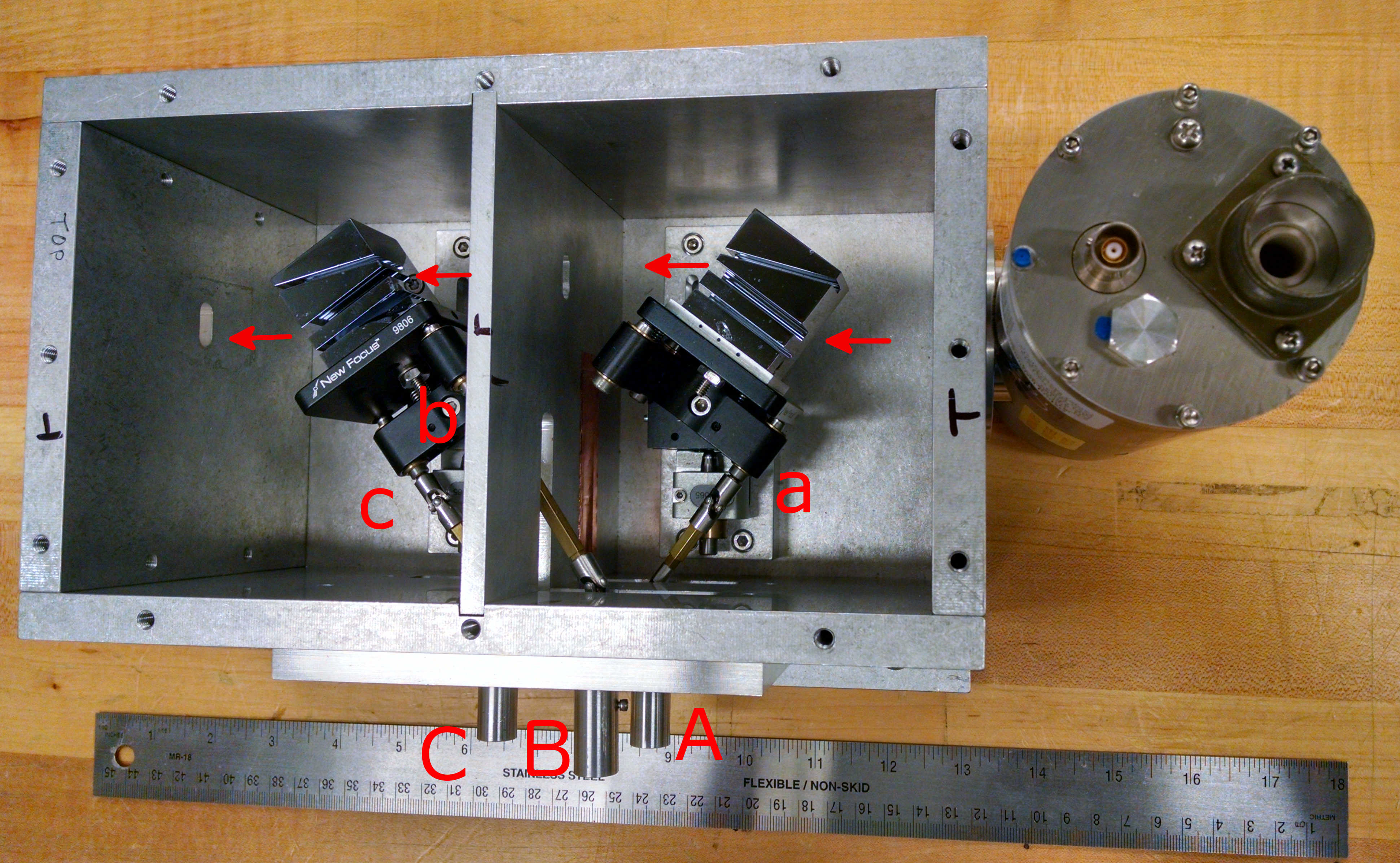}
    \caption{Implementation of \crkaone Si (220) 6+6 reflection monochromator. The top plate has been removed to view the interior. The x-ray tube is the cylinder on the right, and the beam propagates from right to left. The arrows roughly indicate the propagation of the beam outside of the crystal channels. Uppercase labels denote external actuators, while lowercase labels refer to the corresponding adjustors on the optical mounting stage. A: 1st crystal dispersion angle; B: 2nd crystal tip angle; C: 2nd crystal dispersion angle.}
    \label{fig:mono_photo}
\end{figure*}

In Figure~\ref{fig:mono_photo} we show one of our DuMond-Hart-Bartels (DHB) monochromators. A 0.5 in. thick aluminum enclosure provides stable mechanical support for the two channel cut crystals, which are mounted on commercial mirror mount stages, which are in turn mounted on commercial one-dimensional linear positioning stages. An Oxford\footnote{Oxford Instruments X-ray Technology, Scotts Valley, CA 95066, USA} Jupiter 5000 series x-ray tube is mounted at the entrance aperture of the enclosure. Three shafts allow external manipulation of the dispersion angle of each crystal, as well as the relative tip angle of the second crystal. The one-dimensional linear positioning stages are needed only for initial coarse adjustment, and can be accessed by removing the walls of the enclosure from the base plate. The internal plate and back plate absorb any scattered x-rays, allowing photons to travel only along the intended beam path.

\begin{figure}[ht]
    \includegraphics[width=84mm]{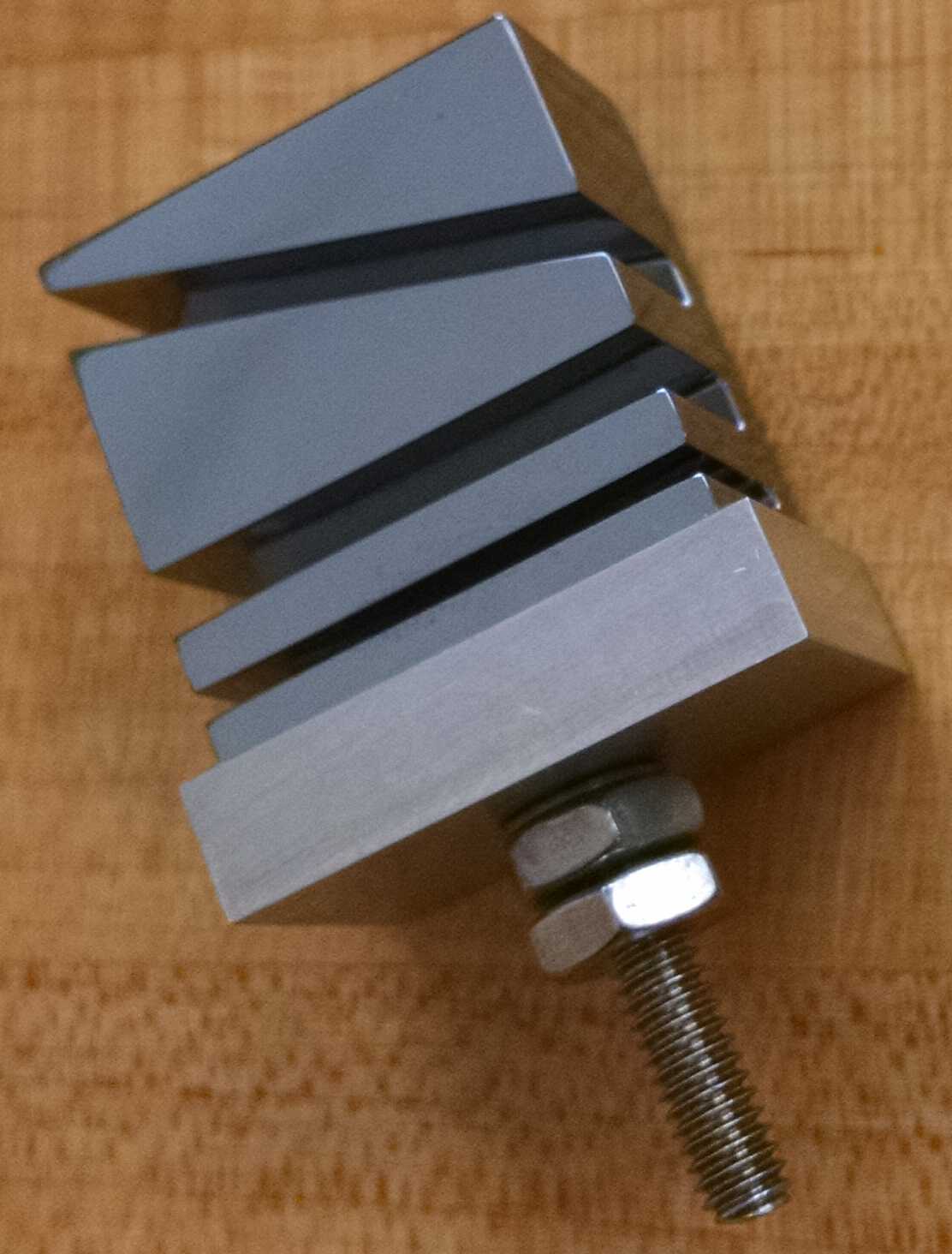}
    \caption{Example of Si (220) channel cut crystal attached to substrate with mounting hardware. The crystal is bonded to the substrate with beeswax. The channel closest to the substrate is designed to prevent propagation of strain from the bond to the reflecting channels. The middle channel is for symmetric Bragg reflection, while the top channel is cut at an angle to the Si (220) plane and can be used for asymmetric Bragg reflection to achieve higher throughput at the price of lower resolution. We only used symmetric reflection in our monochromator.}
    \label{fig:crystal_photo}
\end{figure}

In Figure~\ref{fig:crystal_photo} we show one of the crystals used in our monochromators after it has been bonded to its substrate. The strain relief channel is visible closest to the bond. The crystals are supplied by Crystal Scientific\footnote{Crystal Scientific (UK) Ltd, Middle Barton, Whittingham, Alnwick, Northumberland, NE66 4SU, United Kingdom}, and their manufacturing specifications typically guarantee surface orientation tolerances to within $0.02-0.05^\circ$, spatial tolerances to within $0.05-0.1$ mm, and rocking curve FWHM tolerance to within 0.1 arcsec of the theoretical value.

In Table~\ref{tab:dimensions} we give the dimensions of the crystals used in our monochromators. The ratio of channel width to length is chosen to allow potential use of the full channel width for the design energy using Equation \ref{eqn:crystal_length}, with the exception of the \crkaone 6+6 reflection monochromator, for which we repurposed crystals designed for a 4+4 reflection \cukaone monochromator, taking advantage of the fact that the width-to-length ratio is similar for those two designs, so that almost the full width of the channel is usable. The widths were chosen to allow illumination of a significant part of a calorimeter array with the collimated image (in the dispersion direction) of an x-ray generator filling the full width of the channel; although the x-ray generators used here had much smaller spot sizes (see \S~\ref{sec:beam_width}), in the future we plan to retrofit x-ray generators with larger spot sizes.

\begin{table}[hb]
  \caption{Dimensions of crystals used in monochromators. $l$ gives the length of the channel and $w$ gives the width. \label{tab:dimensions}}
  \begin{ruledtabular}
    \begin{tabular}{cccccc}
      Line & crystal & Number of   & $l$  & $w$ & Bragg angle\\
           &         & reflections & (mm) & (mm)  & (degrees) \\
      \hline \\
      \crkaone & Si (220) & 6+6 & 31.8 & 3.2 & 36.60128\\
      \cukaone & Si (220) & 4+4 & 31.8 & 3.2 & 23.65103 \\
      \cukaone & Si (400) & 4+4 & 52.3 & 9.0 & 34.56447 \\
    \end{tabular}
  \end{ruledtabular}
\end{table}

\subsection{Alignment}

The crystals are aligned using a straightforward procedure. The first crystal (closest to the x-ray tube) is installed and coarsely aligned using a ruler and protractor, while the second crystal is not yet installed. A photon-counting x-ray detector with comparatively large area such as a silicon drift detector or proportional counter is positioned at the expected location of the output beam, coarsely measured using a ruler. The x-ray generator is turned on, and the first crystal reflection angle is varied until the maximum intensity is achieved. Then the generator is turned off, and the second crystal is installed and coarsely aligned. The x-ray detector is placed at the beam exit of the second crystal. The x-ray generator is turned on again, and the output intensity is maximized by varying the second crystal reflection angle, corresponding to the bright \kaone peak of the x-ray generator anode material. This step is very sensitive to the exact angle of the second crystal. While measuring and maximizing the intensity, it should be possible to find a second, smaller (local) maximum corresponding to \katwo at a slightly steeper reflection angle.  Finally, the relative tip angle of the second crystal is adjusted, again by maximizing the intensity. The count rate is much less sensitive to this last degree of freedom, but optimizing the relative tip of the two crystals is important to maximize the monochromaticity of the exit beam along the cross-dispersion direction. 

Alignment of a newly-assembled monochromator typically takes several hours, while realigning a monochromator that is close to the correct alignment takes about an hour. We have verified the short-term alignment stability, for example, before and after positioning the monochromator housing for measurement campaigns. We have not explored the long term alignment stability rigorously, but we believe that realignment is rarely needed, if ever. The maintenance of alignment is verified by remeasuring the count rate of the monochromator for the same x-ray tube settings, and checking that it is consistent with the value obtained during the previous alignment.

\subsection{Mounting}

As will be discussed in \S~\ref{sec:beam_width}, the output beam of the monochromator in the dispersion direction is a collimated image of the x-ray source spot, which is smaller than many calorimeter pixels for typical commercial microfocus x-ray generators, while the beam diverges in the cross-dispersion direction, so that the image on a calorimeter array is a thin line, and only a single column (or row) of such a detector array can be illuminated at one time. To align the monochromator beam to the detector array, and to allow illumination of all pixels, we mounted the entire apparatus (monochromator housing with x-ray tube) on a motorized linear stage. The beam position can then be periodically cycled, allowing illumination of the full array over the course of an experiment.

\section{Theoretical performance \label{sec:theory}}

We used the web tool GID\_SL\cite{1998PhRvB..57.4829S}$^,$\footnote{Please see this URL for access to GID\_SL: \url{http://x-server.gmca.aps.anl.gov/GID_sl.html}} to calculate single reflection rocking curves $R_1(\Delta \theta)$ for relevant x-ray energies and crystal planes. Here $\Delta \theta$ is the offset of the incident ray with respect to the nominal Bragg angle. Rocking curves were calculated for both $\sigma$ and $\pi$ polarization, which had to be considered separately; the results we derive for each polarization are averaged to obtain results for unpolarized light, where appropriate. We assumed symmetric Bragg diffraction from perfect crystals.

\begin{figure}[ht]
  \includegraphics[width=84mm]{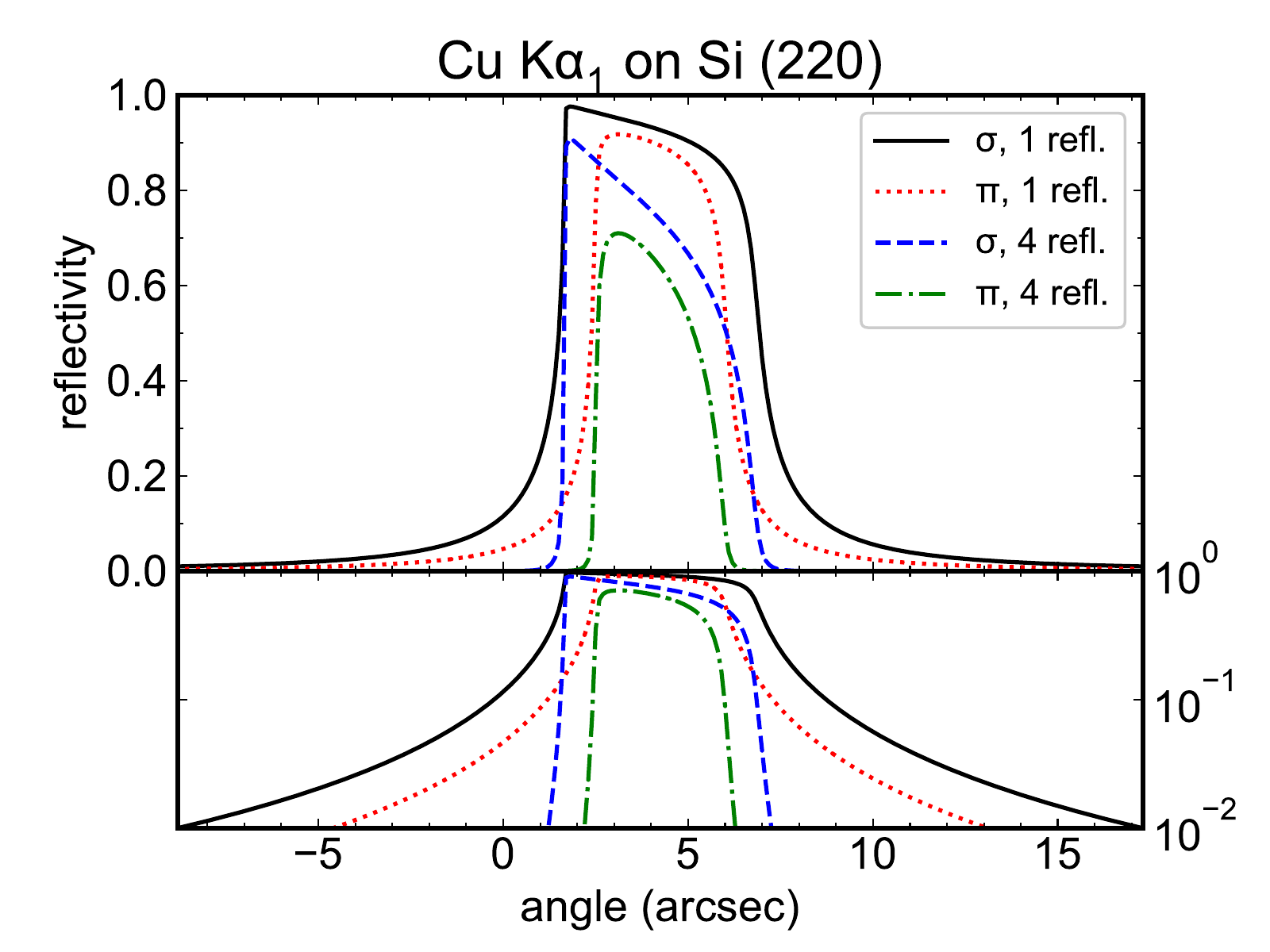}
  \caption{Theoretical reflectivity as a function of angle for the \cukaone peak x-ray energy reflected from Si(220), in both $\sigma$ and $\pi$ polarizations. The top and bottom panels show the same rocking curves with a linear and logarithmic y-axis, respectively. The four-reflection reflectivities are obtained by taking the single-reflection reflectivity to the fourth power. Note that the offset of the rocking curve center of mass on the x-axis is with respect to the nominal Bragg angle for this x-ray energy, and is expected from theory.}
  \label{fig:rock1}
\end{figure}

For each polarization, the rocking curves for $2N$ reflections (with $N$ a positive whole number) from a single channel cut crystal were obtained by multiplying the single crystal rocking curves to the $2N$ power:
\begin{equation}
  R_{2N}(\Delta \theta) = R_1(\Delta \theta)^{2N}\, .
\end{equation}
The rocking curves are shown for both polarizations in Figure~\ref{fig:rock1} for one and four reflections from \cukaone Si~(220).

The rocking curves give the reflectivity as a function of reflection angle with respect to the nominal Bragg angle at the specified energy; note that the actual peak reflectivity is slightly offset from the nominal Bragg angle as expected from theory. This rocking curve is really a one-dimensional version of what can more generally be described as a two-dimensional function of both energy and angle, $R(\Delta E, \Delta \theta)$. In the case of a rocking curve as presented in Figure~\ref{fig:rock1}, the incident energy is fixed and no other energies are considered. To incorporate the effect of other nearby incident energies we displace the rocking curve using a relation derived by dividing the Bragg equation by its derivative:
\begin{equation}
    \frac{\tan \theta}{d\theta} = \frac{\lambda}{d\lambda} = - \frac{E}{dE}\, ,
    \label{eqn:bragg_derivative}
\end{equation} 
so that
\begin{equation}
    d\theta = - \frac{dE}{E}\,\tan\,\theta\, .
\end{equation}
We can thus write the reflectivity as a function of energy and angle in terms of the reflectivity as a function of angle at the nominal energy:
\begin{equation}
    R(\Delta E, \Delta \theta) = R (\Delta \theta^\prime)\, ,
\end{equation}
where
\begin{equation}
    \Delta \theta^\prime = \Delta \theta - \frac{\Delta E}{E} \tan \theta
\end{equation}
is the offset angle with respect to the nominal Bragg angle for a given offset energy $\Delta E$ when evaluated against the geometric reference of the nominal Bragg angle of $\Delta E = 0$.
The reflectivity of the DHB $4N$-reflection monochromator is obtained by multiplying the reflectivity for a single channel cut crystal with $2N$ reflections by itself, but with the second reflectivity having the opposite signs in the displacement of angle as a function of energy displacement:
\begin{equation}
    R_{\mathrm{DHB},4N}(\Delta E, \Delta \theta) = \\R_{2N} (\Delta E, \Delta \theta) R_{2N} (\Delta E, -\Delta \theta)\, .
\end{equation}

To get the transmitted line shape, we integrate the product of the reflectivity and the incident x-ray source flux over incident angles:
\begin{equation}
    F_{E,\mathrm{obs}} (\Delta E)= \int R_{\mathrm{DHB},4N}(\Delta E, \Delta \theta) F_{E,\theta} (\Delta E, \Delta \theta) d\Delta \theta\, ,
\end{equation}
where $F_{E,\mathrm{obs}}$ is the observed flux per unit energy emerging from the monochromator, and $F_{E,\theta}$ is the incident flux per unit energy per unit angle. If we assume that the source emits uniformly over incident angle, as for an X-ray tube, we have
\begin{equation}
    F_{E,\mathrm{obs}} = F_{E,\theta} W_\theta (\Delta E)\, ,
\end{equation}
where
\begin{equation}
    W_\theta (\Delta E) = \int R_{\mathrm{DHB},4N}(\Delta E, \Delta \theta) d\Delta \theta
    \label{eq:angular_acceptance}
\end{equation}
is the {\it angular acceptance}. In the approximation that the incident spectral shape is flat (which is valid for either a continuum, or for a monochromator tuned to the peak of a spectral line that is much broader than the rocking curve), the effective line shape of the monochromator is given by $W_\theta$. Note that $W_\theta (E)$ is effectively a cross-correlation of the rocking curve with its mirror image in $\theta$, with the lag set by the dispersion from the Bragg equation.

\begin{figure}[ht]
  \includegraphics[width=84mm]{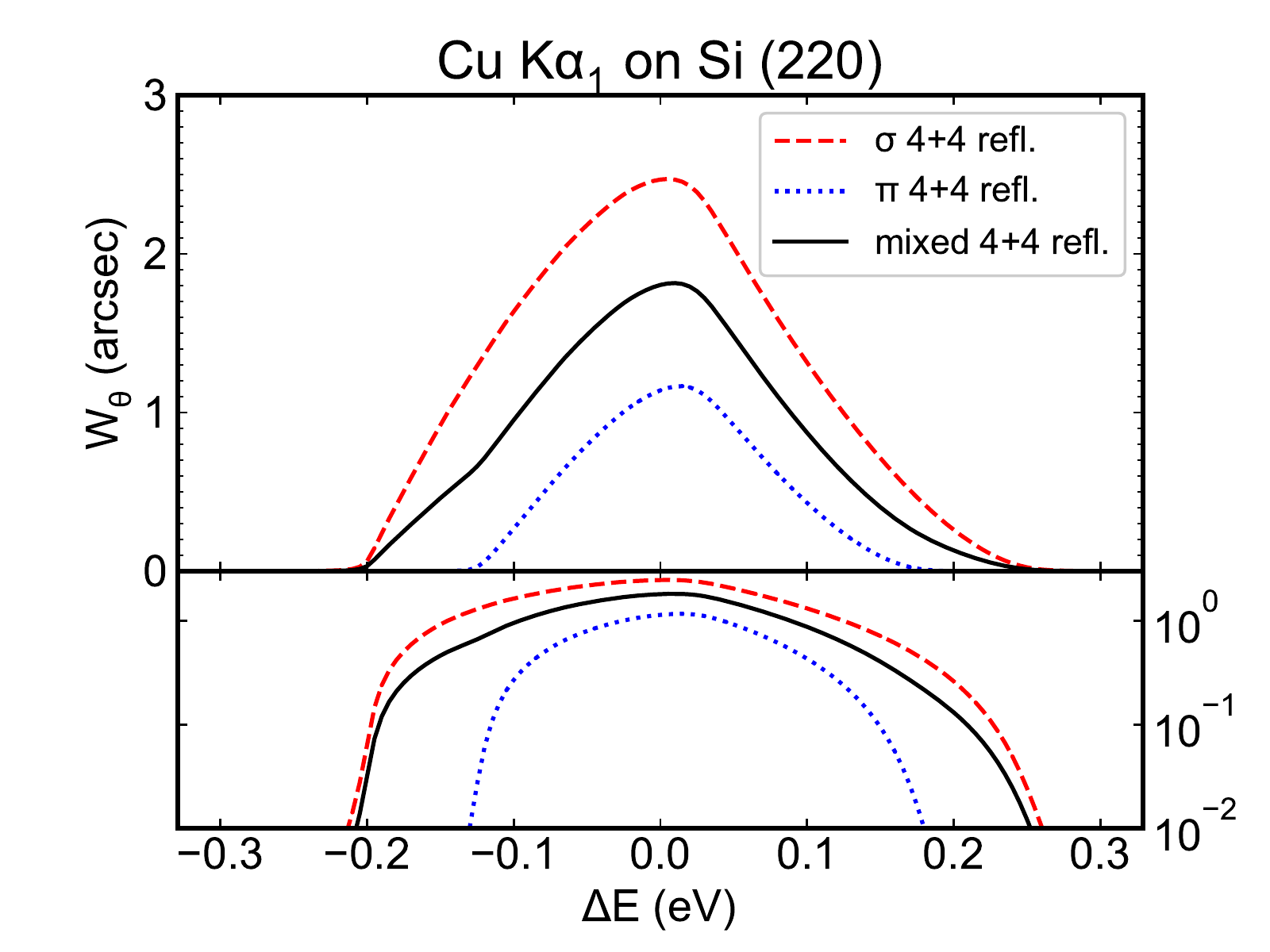}
  \caption{Theoretical angular acceptance as a function of energy for both polarizations of a \cukaone in a 4+4 reflection DHB monochromator using Si (220) crystals, as well as an average for unpolarized light (``mixed''). The top and bottom panels show the same acceptances with a linear and logarithmic y-axis, respectively. These curves were evaluated at the nominal Bragg angle corresponding to the peak energy of \cukaone, but because the center of mass of the rocking curve is slightly offset from the nominal Bragg angle (Figure ~\ref{fig:rock1}), the resulting angular acceptance peak was offset from the nominal energy by 0.35 eV. In practice the true reflection angle is tuned to maximize the flux and therefore the x-axis has been shifted to put the center of mass of the profile at zero. The magnitude of this shift can be used to calculate the expected dispersion angle for optimal alignment of the monochromator; however, in practice it is so small that knowledge of this shift is not useful.}
  \label{fig:rock2}
\end{figure}

\begin{figure}
    \includegraphics[width=84mm]{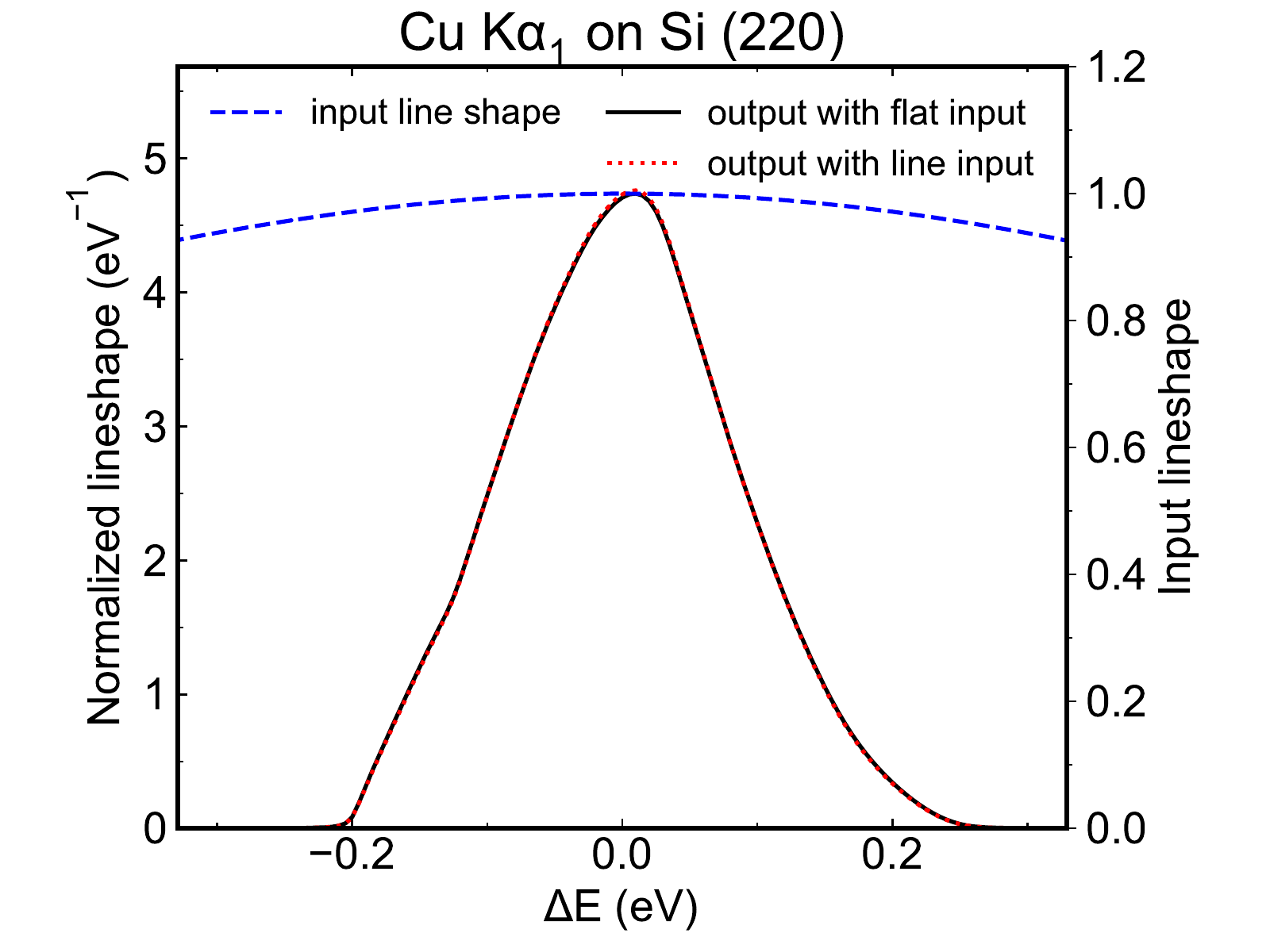}
    \caption{Output line shape for a \cukaone 4+4 reflection DHB monochromator using Si (220) crystals. The black curve shows the line shape for a flat spectral input and has the same shape as the angular acceptance for mixed polarization. The blue dashed curve shows the input line shape from \cukaone. The red dotted curve shows their product, which is the predicted true monochromator line shape. The similarity between the red and black curves shows that it is a good approximation to treat the input as flat, since the monochromator resolution is much higher than the input line width.}
    \label{fig:rock3}
\end{figure}

In Figure~\ref{fig:rock2} we show calculations of $W_\theta$ for a 4+4 reflection Si (220) monochromator for the peak energy of \cukaone. We average $W_\theta$ for $\sigma$ and $\pi$ polarizations to obtain $W_\theta$ for an unpolarized source. In Figure~\ref{fig:rock3}, we show the effect of the \cukaone line shape on the output line shape of the monochromator. Since the input fluorescence line is much broader than the monochromator bandpass, $W_\theta (\Delta E)$ is a good approximation for the true output line shape of the monochromator.

We can also compare the relative efficiency of a DHB monochromator by integrating over both energy and angle to obtain the {\it energy-angular acceptance}, or simply acceptance:
\begin{equation}
    W_{E,\theta} = \int R_{\mathrm{DHB},4N}(\Delta E, \Delta \theta) d\Delta \theta d\Delta E\, .
    \label{eqn:w_e_t}
\end{equation}
Physically, the acceptance accounts for the throughput efficiency of a given energy resolution monochromator due to the rejection of energies outside the bandpass as well as incident angles outside the rocking curve for the accepted energies. The efficiency of a DHB monochromator thus scales with the square of the energy resolution (inversely with the square of the resolving power). 

In Table~\ref{tab:theory} we give the results of our calculations for the FWHM energy resolution and acceptance for the 4+4 \cukaone Si (220) monochromator, as well as for two other configurations we implemented: a 6+6 reflection \crkaone Si(220) monochromator, and a 4+4 reflection \cukaone Si(400) monochromator. We evaluated the formal FWHM energy resolution of the line shape as well as the effective FWHM energy resolution when the line shape is convolved with a 1 eV FWHM Gaussian $\Delta E_{\mathrm{G}}$:
\begin{equation}
    \Delta E_{\mathrm{eff}} = \sqrt{\Delta E_{\mathrm{conv}}^2 - \Delta E_{\mathrm{G}}^2}\, .
\end{equation}
Here $\Delta E_{\mathrm{conv}}$ is the FWHM of the convolution of the monochromator output with $\Delta E_{\mathrm{G}}$. The formal FWHM energy resolution is evaluated according to the formal definition of a full width at half maximum: the values of $\Delta E$ where the profile has half the strength of the peak value are found and then differenced to find the full width. The effective FWHM energy resolution is the value that has to be subtracted in quadrature from an uncorrected calorimeter energy resolution measured with a monochromator in order to obtain the true calorimeter energy resolution. The effective FWHM energy resolution is a figure of merit that more accurately describes the performance of the monochromator in a realistic scenario, such as a measurement of the FWHM energy resolution of calorimeter pixels. The effective FWHM is typically slightly smaller than the formal FWHM. The reason for this difference arises from the fact that convolution is more sensitive to the standard deviation, for which FWHM is not an ideal proxy for a non-Gaussian line shape. We also evaluated the effective FWHM with a range of Gaussian FWHMs, and found that it does not change much as long as the Gaussian FWHM is significantly larger than the monochromator FWHM.

\begin{table}[ht]
  \caption{Predicted energy resolution and acceptance of DHB monochromators. The two values for FWHM energy resolution refer to the formal FWHM of the line shape (form.) and the effective FWHM manifested as an excess broadening when convolved with a 1 eV FWHM Gaussian (eff.). \label{tab:theory}}
  \begin{ruledtabular}
    \begin{tabular}{cccccccc}
      Line & crystal & N. refl. & \multicolumn{2}{c}{$\Delta E$} & \multicolumn{3}{c}{$W_{E,\theta}$}\\
      & & & \multicolumn{2}{c}{(FWHM, eV)} & \multicolumn{3}{c}{(eV arcsec)} \\
      & & & form. & eff. & $\sigma$ & $\pi$ & mixed \\
      \hline \\
      \crkaone & Si (220) & 6+6 & 0.143 & 0.125 & 0.208 & 7.5e-5 & 0.104 \\
      \cukaone & Si (220) & 4+4 & 0.201 & 0.197 & 0.584 & 0.182 & 0.383 \\
      \cukaone & Si (400) & 4+4 & 0.093 & 0.086 & 0.140 & 3.4e-3 & 0.072 \\
    \end{tabular}
  \end{ruledtabular}
\end{table}

\section{Measured performance}

\subsection{Single crystal reflectivity}
\label{sec:single}

To verify the performance of individual crystals, we measured the reflectivity using a collimated monochromatic x-ray source. The source consisted of a water-cooled rotating anode x-ray generator with a Cu target illuminating a single channel cut crystal similar to those used in our DHB monochromators (hereafter referred to as the source crystal, to avoid confusion with the sample crystals being characterized). The x-ray generator was operated with -30 kV cathode bias voltage and 100 mA cathode emission current. We estimate that the x-ray source spot size is $\sim$ 300 $\mu$m. The source crystal was placed at a distance of 2.6 m from the source, with the channel oriented vertically so that the dispersion direction is horizontal. An exit slit with width 100 $\mu$m and height 1 mm was placed at a distance of 0.4~m from the source crystal, and 3~m from the source. The source crystal was configured at an angle of 23.65$^\circ$ relative to the beam path to provide four reflections of \cukaone at 8047.78 eV. The beam divergence is set by the source spot size and exit slit width in comparison with their 3~m separation. Adding the source spot size and exit slit width in quadrature, we estimate an effective slit size of 316 $\mu$m and thus a divergence in the dispersion direction of 0.105 milliradian. Using Equation~\ref{eqn:bragg_derivative}, we find that this corresponds to an energy width of 1.93 eV. This is comparable to the FWHM of \cukaone, so we expect both the natural line shape and the slit width to contribute to the source line shape.

The sample crystal was mounted on a goniometer allowing rotation of the crystal along the dispersion direction, and the reflected x-rays were measured with a Xe-filled proportional counter. The distance from the source exit slit to the goniometer was 25 cm, and the distance from the goniometer to the proportional counter was 25 cm. The raw flux of \cukaone passing through the slit was measured by moving the sample out of the beam path and moving the proportional counter to the location of the unreflected beam. The experimental setup is the same as described in Figure 2 of \citet{10.1117/12.2313469}, with the exception that the Si-PIN diode x-ray detector was replaced with the proportional counter, and the x-ray detector was not positioned at twice the goniometer angle, but rather at the correct position to intercept the reflected beam from the sample crystal, which is displaced parallel to the incident beam path.

\begin{figure}[ht]
  \includegraphics[width=84mm]{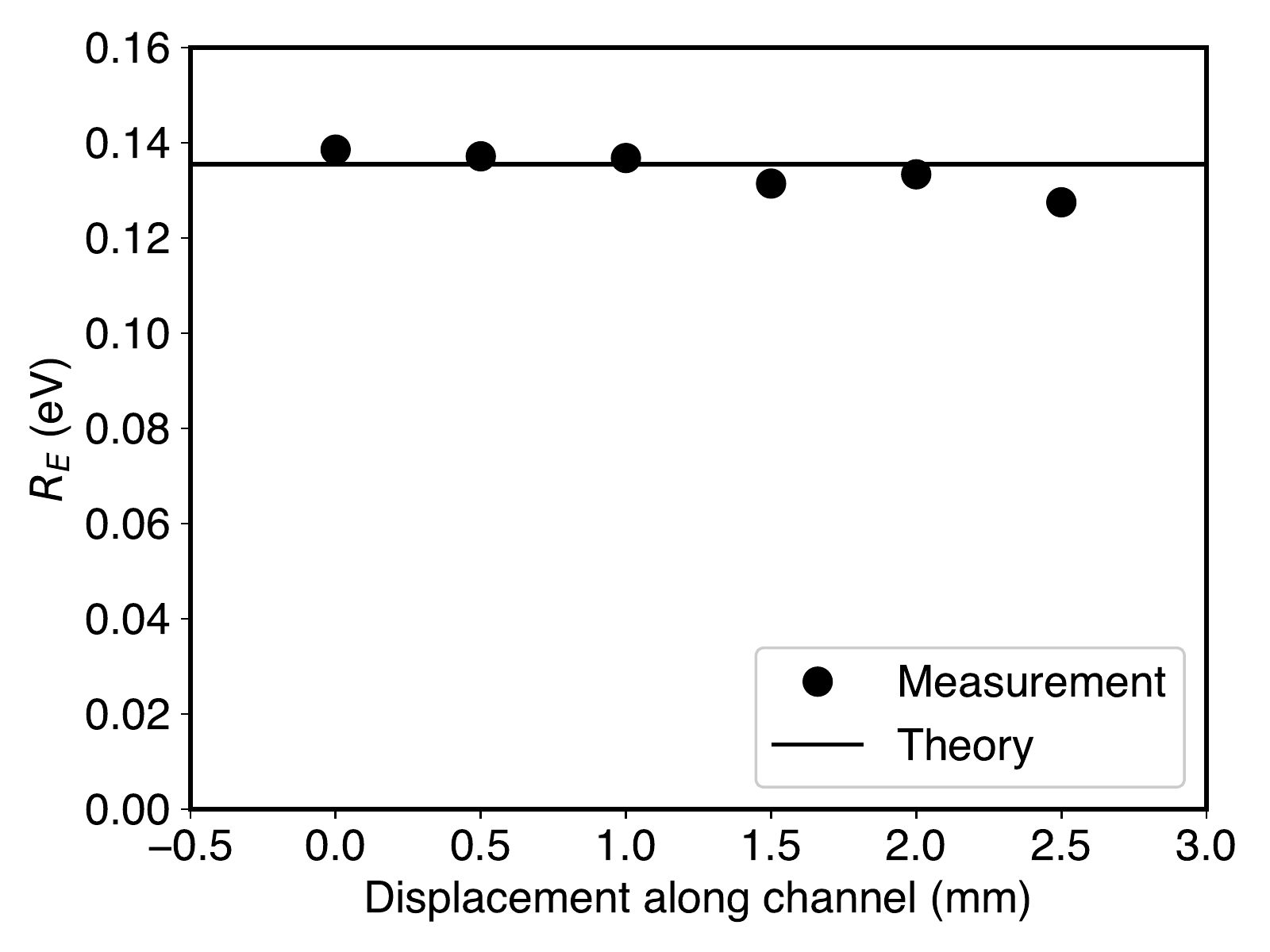}
  \caption{Measured energy-integrated reflectivity ratio $\mathcal{R}_E$ as a function of displacement along the channel width for a Si (220) crystal reflecting monochromatic \cukaone photons. The predicted $R_E$ is shown as a solid horizontal line. This crystal was bonded to the substrate with epoxy.}
  \label{fig:EW_Si220_old}
\end{figure}

\begin{figure}[ht]
  \includegraphics[width=84mm]{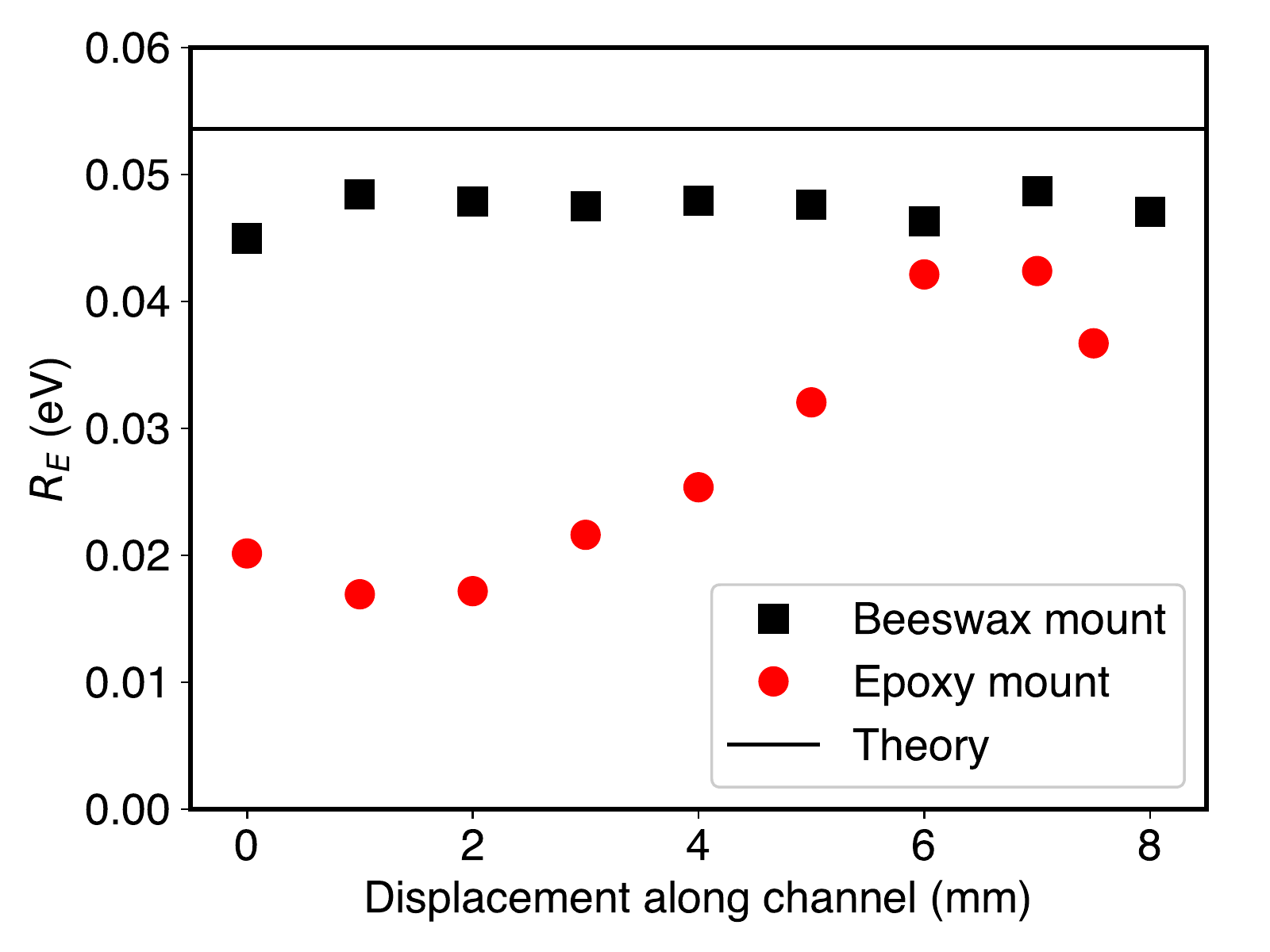}
  \caption{Measured energy-integrated reflectivity ratio $\mathcal{R}_E$ as a function of displacement along the channel width for a Si (400) crystal reflecting monochromatic \cukaone photons. This crystal was first bonded to the substrate with epoxy (red circles). Because the reflectivity was poor, the epoxy was removed and the crystal was rebonded with beeswax (black squares), leading to better reflectivity more closely matching the theoretical prediction.}
  \label{fig:EW_Si400}
\end{figure}

We measured the reflected flux as a function of goniometer angle and displacement of the sample crystal along the channel direction. To understand how to interpret these measurements, we need to further develop the theory discussed in \S~\ref{sec:theory}. The source and sample crystal act effectively as a DHB monochromator; however, we must now account for the fact that the source and sample crystals may be different. We still need to satisfy the Bragg equation for both crystals, so following Equation~\ref{eqn:bragg_derivative}, we have
\begin{equation}
\frac{-\Delta E}{E} = \frac{\Delta\lambda}{\lambda} = \frac{\Delta\theta_A}{\tan \theta_A} = \frac{\Delta\theta_B}{\tan \theta_B}\, .
\end{equation}
Here we use subscripts A and B to refer to the source and sample crystals, respectively. The change in the goniometer angle is the sum of the change in the two crystal reflection angles: $\Delta \theta_g = \Delta \theta_A + \Delta \theta_B$. Thus
\begin{equation}
    \Delta \theta_g = \frac{\Delta E}{E} (\tan\theta_A + \tan\theta_B)\, .
\end{equation}

If we denote the input flux from the x-ray generator as a function of energy and angle as $F_{E,\theta}$, and if we again assume that the input line width is much broader than the line width passed by the two crystals, then the flux observed as a function of goniometer angle is 
\begin{equation}
F^{\mathrm{AB}}_{\mathrm{obs}} (E) = F_{E,\theta} (E) W_{E,\theta}^{\mathrm{AB}}\, , 
\label{eq:f_obs_refl}
\end{equation}
where the superscript AB indicates that the acceptance is evaluated for the reflectometer two-crystal system of the source and sample crystals, and where the energy $E$ is that allowed by the choice of goniometer angle. Here $W_{E,\theta}^{\mathrm{AB}}$ is calculated as in Equation~\ref{eqn:w_e_t}, but accounting for the heterogeneity of the source and sample crystal:
\begin{equation}
    W_{E,\theta}^{\mathrm{AB}} = \int R^\mathrm{A} (\Delta E, \Delta \theta_A) R^\mathrm{B} (\Delta E, \Delta \theta_B) d\Delta \theta_g d\Delta E\, .
    \label{eqn:w_e_t_AB}
\end{equation}
Here we use the abbreviated notation $R^i$, with $i=A,B$, to refer to $R_{2N}$ for the source and sample crystals, respectively. Note that near line center, the slit setting is irrelevant, as the bandpass of the source-sample crystal system is much narrower than the slit width. However, as the goniometer is scanned, the slits do set the bounds of the scan angle where significant flux can be detected.

The flux collected by the x-ray detector with no sample crystal present is given by
\begin{equation}
    F_{\mathrm{obs}} = \int^{\Delta\theta_+}_{\Delta\theta_-}\int^{\infty}_{0} F_{E,\theta} R^\mathrm{A} (\Delta E, \Delta \theta_A) dE d\Delta \theta_A\, .
\end{equation}
The angular integration limits are set by the effective slit width.
In analogy with Eq.~\ref{eq:angular_acceptance}, we define the {\it non-dispersive} angular acceptance
\begin{equation}
    W_{2N,\theta} \equiv \int R_{2N} (E, \Delta \theta) d\Delta \theta\, .
\end{equation}
Note that in principle $W_{2N, \theta}$ is still a function of energy, but since the rocking curve shape does not change much for a small change in energy, it is effectively a constant. Again using A to refer to the source crystal, we then have
\begin{equation}
    F^{\mathrm{A}}_{\mathrm{obs}} = W_{2N,\theta}^{\mathrm{A}} \int^{\Delta E_+}_{\Delta E_-} F_{E,\theta} dE\, ,
    \label{eq:f_obs_source}
\end{equation}
where we have converted the angular integration limits into equivalent energy integration limits. We have again assumed a uniform angular distribution for emission from the source.

Using Eqs.~\ref{eq:f_obs_refl} and \ref{eq:f_obs_source}, the ratio of the flux observed with the sample crystal in and out is
\begin{equation}
\mathcal{R} = \frac{W_{E,\theta}^{\mathrm{AB}}}{W_{2N,\theta}^\mathrm{A}} \frac{F_{E,\theta}}{\int F_{E,\theta} dE}\, .
\end{equation}
This is still not useful without knowledge of the line shape from the x-ray generator. Since we can scan the goniometer angle with the sample crystal installed, which is equivalent to scanning the energy passed by the system, we can thus effectively integrate the observed ratio over energy:
\begin{equation}
    \mathcal{R}_E = \int \mathcal{R} dE = \frac{W_{E,\theta}^{\mathrm{AB}}}{W_{2N,\theta}^\mathrm{A}}\, .
\end{equation}
Because the same integration limits imposed by the slits apply to both integrals over the source spectrum, the integrated spectrum terms cancel. Finally, note that for both $W_{E,\theta}^{\mathrm{AB}}$ and $W_{2N,\theta}^\mathrm{A}$, we must again evaluate them for both polarizations and average them to get the unpolarized values. We use the rocking curves discussed in Section \ref{sec:theory} to calculate theoretical values of $\mathcal{R}_E$ for comparison to our measurement data. 

Results representative of our measurements are shown in Figures~\ref{fig:EW_Si220_old} and \ref{fig:EW_Si400}. The first pair of crystals we used were Si (220) crystals with a $\sim 3$ mm wide channel. These were bonded to Al substrates with Henkel\footnote{Henkel Corporation Aerospace, 2850 Willow Pass Road, Bay Point, CA 94565, USA} Loctite Hysol 9309 epoxy. We found that the reflectivity matches the prediction (Fig.~\ref{fig:EW_Si220_old}). We repeated the procedure for a pair of Si (400) crystals with a $\sim 9$ mm wide channel. We found that the reflectivity was significantly less than the prediction. We suspected strain due to the epoxy bond was responsible, so we removed the epoxy and rebonded the crystals to the substrates using beeswax. We remeasured the reflectivity and found that it was only slightly less than predicted (Fig.~\ref{fig:EW_Si400}). It is not clear why the Si (400) crystals showed strain when bonded with epoxy while the Si (220) crystals did not. This difference may be related to the relative width of the channels or to details of the epoxy bonding procedure for each crystal pair; however, beeswax gives good results and is much easier to unbond and rebond, so we recommend its use for the current application.

We note that the reduced reflectivity of the strained crystals would significantly reduce the output flux from the full monochromator. It is thus essential to achieve a sufficiently strain-free bond between the crystals and their substrates, and prudent to validate the reflectivity of individual crystals before assembling the full monochromator.

\subsection{Performance of full monochromator}

We tested three different monochromator configurations and measured the count rate in each configuration incident on an Amptek\footnote{Amptek Inc., 14 DeAngelo Drive, Bedford, MA 01730, USA} XR-100CR Si-PIN detector with 13 mm$^2$ area positioned at the exit aperture of the monochromator box. We list the count rates in Table~\ref{tab:countrate}. We also give count rates per angle in the cross dispersion direction, using an estimated cross dispersion capture angle of 16.0 mrad for our experiment configuration. The beam is collimated in the dispersion direction.

We note that the ratio of theoretical $W_{E,\theta}$ for the two \cukaone monochromators is 5.3, while the ratio of count rates observed is 12.0 (after correcting for the different emission current in the two experiments). It is not clear where this factor-of-two discrepancy in efficiency originates; the reflectivity reduction below theory shown in Figure~\ref{fig:EW_Si400} is about 10\%, so that a 20\% reduction in efficiency with respect to theory would be expected for the full monochromator.

\begin{table}[ht]
  \caption{Measured count rates in test configuration \label{tab:countrate}. $V$ is the x-ray tube anode bias voltage, and $I_{em}$ is the cathode emission current.}
  \begin{ruledtabular}
    \begin{tabular}{ccccccc}
      \multicolumn{3}{c}{configuration} & rate & rate & $V$ & $I_{em}$ \\
      Line & crystal &  reflections & (c/s) & (c/s/mrad) & (kV) & (mA) \\
      \hline \\
      \crkaone & Si (220) & 6+6 & 179 & 11.2 & 15 & 1 \\
      \cukaone & Si (220) & 4+4 & 921 & 57.6 & 20 & 0.7 \\
      \cukaone & Si (400) & 4+4 & 110 & 6.9 & 20 & 1 \\
    \end{tabular}
  \end{ruledtabular}
\end{table}

\subsubsection{Beam width}
\label{sec:beam_width}

\begin{figure}[ht]
  \includegraphics[width=84mm]{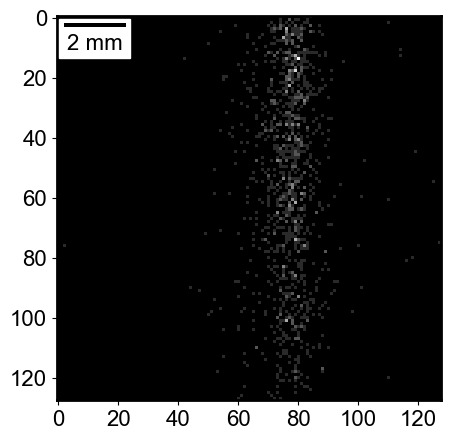}
  \caption{CCD image of monochromator beam. The native pixel size of the CCD is 24 $\mu$m, and the pixels have been rebinned by a factor of four on both axes in this image.}
  \label{fig:ccd_image}
\end{figure}

\begin{figure}[ht]
  \includegraphics[width=84mm]{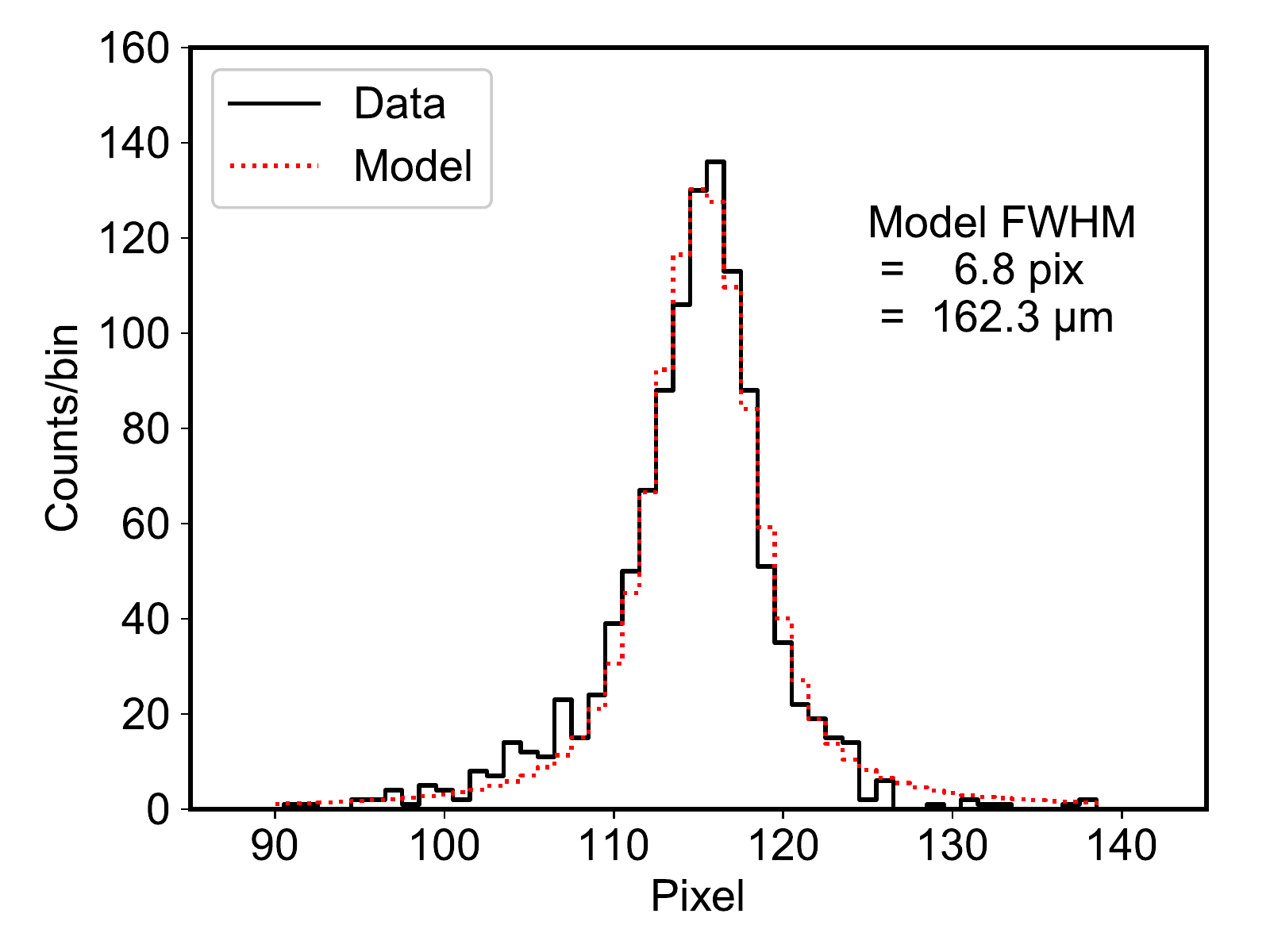}
  \caption{Histogram in the dispersion direction of CCD image of monochromator beam. The pixel size is 24 $\mu$m, and there has been no rebinning. The image was rotated before histogram accumulation to correct for a slight misalignment of the beam direction to the native pixel grid. The best fit Voigt function has a FWHM of 6.8 pixels, corresponding to 162.3 $\mu$m.}
  \label{fig:ccd_image_histo}
\end{figure}

The optical design of the monochromator produces a collimated image of the anode spot in the dispersion direction, and a radially diverging beam in the cross-dispersion direction. To characterize the beam width in the dispersion direction, we imaged the \crkaone Si (220) monochromator with a Princeton PI-SX:512 CCD. The x-ray tube was operated with anode bias voltage $V = 10 $~kV and emission current $I_{em} = 0.5$~mA. The CCD has 24 $\mu$m $\times$ 24 $\mu$m pixels, allowing measurement of the beam size with good resolution. It was operated at 1 MHz readout rate and with an exposure time of 1 s, so that individual x-ray events could be counted, resulting in a frame time of 1.26 s and an out-of-time event fraction of 21\%. The frame transfer direction was oriented parallel to the cross-dispersion axis of the monochromator, so that the frame streak from out-of-time events runs in the same direction as the source image. Cosmic ray events were identified and removed based on both pulse height and charge distribution pattern. The image is shown in Figure~\ref{fig:ccd_image}, and a histogram of the image in the dispersion direction is shown in in Figure~\ref{fig:ccd_image_histo}. We fit the histogram with a Voigt function, obtaining a FWHM of 6.8 pixels, or 162 $\mu$m.

\begin{figure*}[ht]
    \includegraphics[width=168mm]{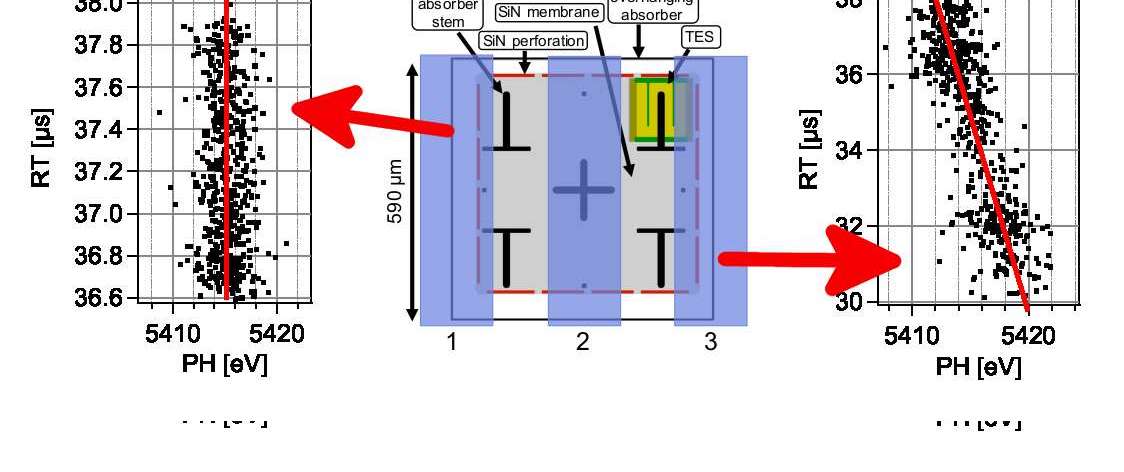}
    \caption{Test of position dependent response in TES detector \citep{2013ASC_Eckart}. The central diagram shows three beam positions where pulses were recorded. The left and right hand scatter plots show pulse risetime (RT) vs. pulse height (PH) for events recorded at beam positions 1 and 3, respectively. The events in position 3 show a clear RT-PH correlation, which is due to a dependence of RT and PH on the distance from the TES sensor to where the photon was absorbed. To achieve the required energy resolution for this application, absorption of an x-ray with a given energy should result in a similar pulse shape independent of where it arrives on the area of the absorber. The observed RT-PH correlation and its variation with photon beam position on the absorber indicates problems with the thermalization of incident photons in the absorber. The results of this experiment were used to identify the origin of a measured distortion in the spectral response of these detectors that could then be addressed by changes to the pixel design and fabrication processes. Part of this figure is adapted with permission from IEEE Transactions on Applied Superconductivity 23, 2101705 (2013). Copyright 2013 IEEE.}
  \label{fig:microX}
\end{figure*}

The manufacturer reported a measurement for the anode spot size of the Cr anode x-ray generator (serial number 59695) of $114.3 \pm 1.2 \mu$m FWHM in the direction we aligned with the dispersion axis of the monochromator. The manufacturer's measurement was conducted at operating conditions of $V = 50$ kV and $I_{em} = 1$ mA. The anode spot size is expected to change somewhat as a function of x-ray tube operating parameters, with spot size increasing with increased emission current, and decreasing as anode voltage increases. The FWHM of the DHB rocking curve for \crkaone on Si (220) is 8 arcsec, which corresponds to 24 $\mu$m at the measurement distance of 60 cm between the crystal and CCD. The correction due to angular divergence is thus almost negligible when added in quadrature with the anode spot size. We conclude that the measured beam width is consistent with the expectation of a collimated image of the anode spot, given the systematic uncertainty associated with the different tube operating conditions.

A beam width of 160 $\mu$m is smaller than the pixels of many calorimeter arrays, although arrays with smaller pixels have been fabricated. One possible application of a narrow beam monochromator is in tests for an undesired position dependent response. One example of this is shown in \citet{2013ASC_Eckart}, who used the narrow beam of the monochromator to probe position dependent response in a TES array developed for the Micro-X sounding rocket mission\cite{Adams2020}. The results of this test are summarized in Figure~\ref{fig:microX}.

\subsubsection{Energy resolution}

It is difficult to directly measure the energy resolution of a DHB monochromator of the design discussed in this article, since the output x-ray flux is low by design, and furthermore the resolving power of a spectrometer used to measure the energy resolution of the monochromator must be comparable to the monochromator itself.

\begin{figure*}[ht]
  \includegraphics[width=168mm]{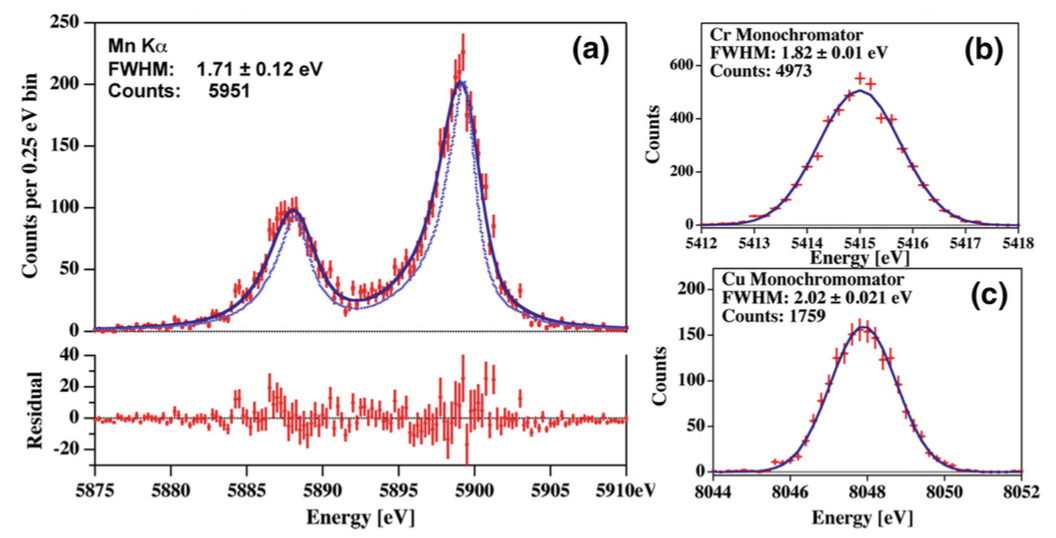}
  \caption{Measurements of the energy resolution of a metallic magnetic x-ray calorimeter from \citet{2014JLTP..176..617P}. Panel (a) shows a measurement of the Mn K$\alpha$ line from a radioactive $^{55}$Fe source; the quoted energy resolution is the modelled intrinsic detector resolution after accounting for the known line shape of Mn K$\alpha$. Panels (b) and (c) show measurements using the 6+6 reflection Si (220) Cr K$\alpha$ monochromator and 4+4 reflection Si (400) Cu K$\alpha$ monochromator, respectively; the quoted energy resolution for these models does {\it not} account for the broadening from the monochromator. Reproduced with permission from Journal of Low Temperature Physics 176, 617 (2013). Copyright 2013 Springer Nature.}
  \label{fig:magcal}
\end{figure*}

\citet{2014JLTP..176..617P} have measured the energy resolution of an MMC using an \fefiftyfive source, as well as our portable DHB monochromators (reproduced in Figure~\ref{fig:magcal}). We can use this to estimate experimental constraints on the resolution of the \crkaone Si (220) monochromator, since the resolution at \mnka and \crkaone should be similar. The measured detector energy resolution from \mnka after deconvolving the natural line shape\cite{1997PhRvA..56.4554H} is 1.71 $\pm$ 0.12 eV, while  the Gaussian FWHM when illuminated by the \crkaone monochromator is 1.82 $\pm$ 0.01 eV, with no deconvolution. If we correct to this by subtracting the theoretical 0.125 eV FWHM of the monochromator in quadrature, we obtain 1.81 $\pm$ 0.01 eV. In either case, the resolution is consistent within 1$\sigma$ with that measured using the \fefiftyfive source. Formally, the measured energy resolution of the monochromator obtained by differencing the \fefiftyfive and \crkaone measurements in quadrature and propagating the measurement uncertainty is $\Delta E_{\mathrm{FWHM}}$ = 0.6 $\pm$ 0.3 eV. However, this is of questionable validity, since the propagated uncertainty on the energy resolution depends inversely on the best fit energy resolution itself, which is clearly very uncertain. If we instead neglect the relatively small uncertainty on the measurement with the monochromator, and use the 1$\sigma$ upper and lower bounds of the \mnka measurement to derive the bounds on the monochromator resolution, we obtain an upper limit of 0.9 eV FWHM. This is consistent with the theoretical resolution of 0.125 eV. We cannot use the MMC measurements presented in  Figure \ref{fig:magcal}c using the \cukaone Si (400) monochromator in a similar way, both because the energy resolution at \cukaone is expected to be different from that at \mnka, and because the detector operating conditions had changed for this exposure due to a difference in the laboratory noise environment.

\begin{figure*}[ht]
  \includegraphics[width=168mm]{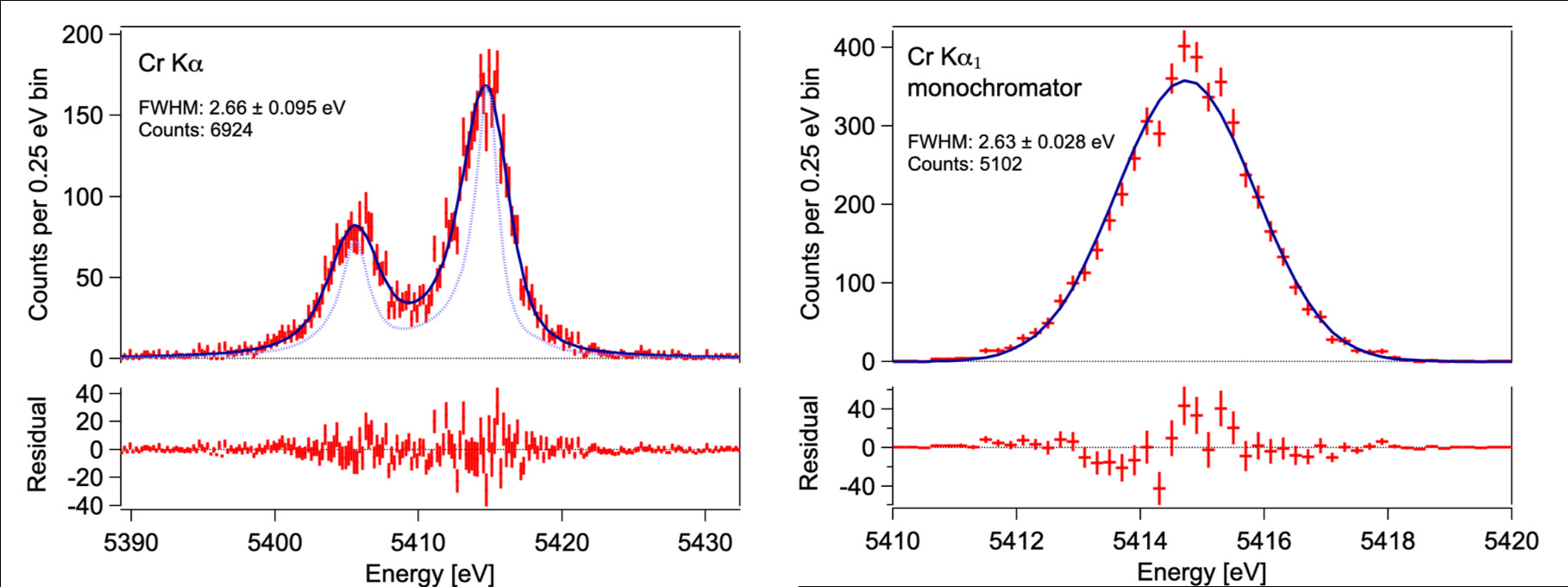}
  \caption{X-ray spectra measured with a TES microcalorimeter. Illumination with \crka fluorescence (left) and the \crkaone monochromatic photons (right). The labeled Gaussian FWHM energy resolution for the \crka complex represents the intrinsic detector resolution after deconvolving the natural line shape\cite{1997PhRvA..56.4554H}, whereas the FWHM measured with the monochromator is reported with no deconvolution.}
  \label{fig:TESspectra_doublewide}
\end{figure*}

We also used a TES microcalorimeter to measure the energy resolution of the monochromator. The detector was from a prototype Large Pixel Array\cite{Smith2016XIFU} for the Athena X-ray Integral Field Unit (X-IFU)\cite{2018SPIE10699E..1GB, 2018cosp...42E2549P}. We obtained x-ray spectra from a \crka fluorescent source and the \crkaone monochromator, providing a comparison at the same incident energy. The measurements were performed consecutively on the same day with identical detector and cryogenic operating condition setpoints. The noise environment was stable. The results are presented in Figure~\ref{fig:TESspectra_doublewide}. Just as in our measurements using the MMC, we find that these two measurements are consistent at the 1$\sigma$ level, and the inferred upper limit to the FWHM energy resolution of the monochromator is 0.7 eV, consistent with the theoretical resolution of 0.125 eV.

In both of the experiments described in this section, the reported error estimates are derived only from propagation of statistical uncertainties in fits to the data. The measurements are fully consistent with the theoretical prediction within the uncertainties. A more stringent measurement of the monochromator energy resolution is desirable, given ever-improving calorimeter energy resolutions, but will be challenging to achieve.

\subsubsection{Polarization}
\label{sec:polarization}

The Cr \kaone 6+6 reflection Si 220 monochromator is theoretically predicted to be 99.96\% $\sigma$-polarized, i.e., with the electric field in the dispersion direction. This makes it potentially of interest as a monochromatic calibration source for x-ray polarimeters.

\begin{figure}[ht]
  \includegraphics[width=84mm]{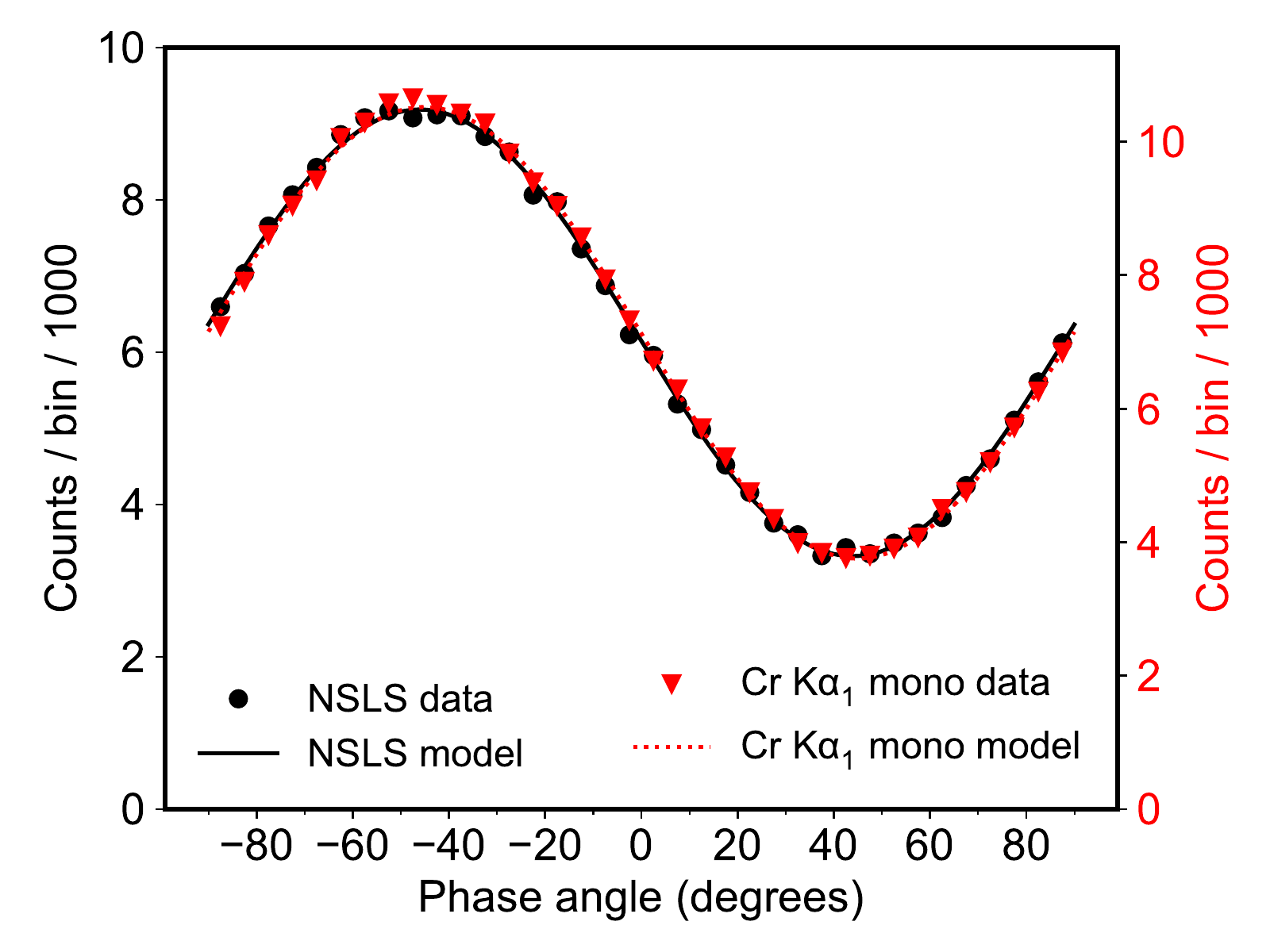}
  \caption{Polarization measurements using the TPC polarimeter for the \crkaone Si (220) 6+6 reflection monochromator compared with calibration measurements from beamline X-19A at the NSLS synchrotron facility. The x-axis denotes the reconstructed angle of the track of the initial photoelectron, while the y-axis gives the number of counts accumulated per 5 degree angle bin. The left (black labels) and right (red labels) y-axes are for the NSLS and \crkaone monochromator data and models, respectively. The solid lines show the best fit sine plus constant models. The modulation factor is derived by dividing the amplitude of the sine by the constant y-axis offset, and is proportional to polarization.}
  \label{fig:polarization}
\end{figure}

We measured the polarization of this monochromator to be 99 $\pm$ 1 \% (statistical) using a micropattern Time Projection Chamber (TPC) polarimeter \cite{2016NIMPA.838...89I}, which obtains polarization information by analyzing the angle of the track created by the initial photoelectron. The polarimeter was calibrated using beamline X19-A at the National Synchrotron Light Source (NSLS), which has a calculated polarization of 98\%. In Figure~\ref{fig:polarization} we show the number of events collected per phase angle bin in the TPC for both the synchrotron calibration and our measurement of the Cr \kaone 6+6 reflection Si 220 monochromator. The data are modeled with a sine plus a constant baseline. The ratio of the sine amplitude to the constant baseline is known as the modulation factor, and is proportional to the degree of polarization.

The polarization measurement of the monochromator may have some additional experimental errors that we estimate to be comparable to or smaller than the statistical uncertainty. The response of the polarimeter varies both with energy and, to a lesser extent, with position in the active area.  We believe that the energy and position of the beams for the two measurements were sufficiently similar that these errors are small compared to the statistical uncertainties. There may also be an error in the assumed synchrotron polarization of 98\%, which is derived from a ray-tracing calculation\cite{1990NIMPA.291..157Y}, but has not been experimentally verified. Regardless of any possible systematic errors, our measurements are compatible with the theoretical prediction at the 1$\sigma$ level.

\section{Summary}

We have reported on the design and construction of simple, portable x-ray monochromators intended for calibration of high-resolution x-ray calorimeters. The typical FWHM energy resolution of our designs is 0.1-0.2 eV, and the output count rate may be as high as 500 counts/s/mm$^{2}$ at the monochromator exit aperture for typical operating conditions. The illumination pattern is a narrow strip with width corresponding to the x-ray tube spot size, $\sim$ 100-200 $\mu$m for the monochromators we have implemented.

Future work includes the development of further monochromators for different fixed energies, including Ti K$\alpha_1$, Fe K$\alpha_1$, and Au L$\alpha_1$ and $\beta_1$; implementation of computer-controlled, motorized positioning for the crystals; and use of bright, water cooled x-ray generators with large anode spots, together with broad channels cut in the crystals. The latter is particularly important for the Athena X-IFU and other instruments featuring large arrays of calorimeter pixels, so that much or all of the array may be simultaneously illuminated, allowing for more efficient use of limited calibration time. Many of these features have been implemented and are undergoing performance testing in our laboratory.

\begin{acknowledgments}
We thank two anonymous referees for providing insightful comments that significantly helped the clarity and presentation of this article.
We thank T. Okajima for use of the x-ray reflectometer, and members of the GSFC x-ray calorimeter research groups for the results presented in Figures 12-14. This work was supported by NASA's Astrophysics Division. Part of this work was performed under the auspices of the U.S. Department of Energy by Lawrence Livermore National Laboratory under Contract DE-AC52-07NA27344. 
\end{acknowledgments}

\section*{Data Availability Statement}
The data that support the findings of this study are available from the corresponding author upon reasonable request.

\bibliography{master.bib}

\end{document}